%% file: main.tex
\documentclass[default]{sn-jnl}

\usepackage{graphicx} 
\usepackage{amsmath,amsfonts}
\usepackage{textcmds}
\usepackage{forest}
\usepackage{flexisym}
\usepackage[utf8]{inputenc}
\usepackage{longtable}
\usepackage{multirow}
\usepackage{manyfoot}
\usepackage{pifont}
\usepackage{pdflscape}
\usepackage{tikz,xcolor}
\usepackage{url}
\usepackage{orcidlink}
	\definecolor{bleudefrance}{rgb}{0.19, 0.55, 0.91}
	\definecolor{britishracinggreen}{rgb}{0.0, 0.26, 0.15}

\usetikzlibrary{arrows.meta, shapes.geometric, calc, shadows}
\usetikzlibrary{arrows,shapes,positioning,shadows,trees}
\usetikzlibrary{chains,positioning,shapes.symbols,fadings,shadows, backgrounds}
\usetikzlibrary{decorations.pathmorphing}
\usetikzlibrary{shapes.callouts}
\usetikzlibrary{shapes.arrows, shadings}
\usetikzlibrary{decorations.text}
\usetikzlibrary{shapes.geometric, arrows}

\colorlet{mygreen}{green!75!black}
\colorlet{col1in}{red!30}
\colorlet{col1out}{red!40}
\colorlet{col2in}{mygreen!40}
\colorlet{col2out}{mygreen!50}
\colorlet{col3in}{blue!30}
\colorlet{col3out}{blue!40}
\colorlet{col4in}{mygreen!20}
\colorlet{col4out}{mygreen!30}
\colorlet{col5in}{blue!10}
\colorlet{col5out}{blue!20}
\colorlet{col6in}{blue!20}
\colorlet{col6out}{blue!30}
\colorlet{col7out}{orange}
\colorlet{col7in}{orange!50}
\colorlet{col8out}{orange!40}
\colorlet{col8in}{orange!20}
\colorlet{linecol}{blue!60}

\usepackage{colortbl}
\definecolor{LightCyan}{rgb}{0.88,1,1}
\definecolor{darkgreen}{rgb}{0.0, 0.2, 0.13}
\definecolor{darkpastelgreen}{rgb}{0.01, 0.75, 0.24}

\forestset{qtree/.style={for tree={parent anchor=south, 
			child anchor=north,align=center,inner sep=0pt}}}
 
\tikzstyle{startstop} = [rectangle, rounded corners, minimum width=3cm, minimum height=1cm,text centered, draw=black, fill=white!30]
\tikzstyle{arrow} = [thick,->,>=stealth]
\def\tsc#1{\csdef{#1}{\textsc{\lowercase{#1}}\xspace}}
\tsc{WGM}
\tsc{QE}
\tsc{EP}
\tsc{PMS}
\tsc{BEC}
\tsc{DE}

\definecolor{lime}{HTML}{A6CE39}
\DeclareRobustCommand{\orcidicon}{%
	\begin{tikzpicture}
	\draw[lime, fill=lime] (0,0) 
	circle [radius=0.16] 
	node[white] {{\fontfamily{qag}\selectfont \tiny ID}};
	\draw[white, fill=white] (-0.0625,0.095) 
	circle [radius=0.007];
	\end{tikzpicture}
	\hspace{-2mm}
}

\foreach \x in {A, ..., Z}{%
	\expandafter\xdef\csname orcid\x\endcsname{\noexpand\href{https://orcid.org/\csname orcidauthor\x\endcsname}{\noexpand\orcidicon}}
}

\begin{document}

\title[Demystifying Object-based Big Data Storage Systems]{Demystifying Object-based Big Data Storage Systems}

\author*[1]{\fnm{Anindita} \sur{Sarkar Mondal}\orcidA}
\email{sarkar.anindita5@gmail.com}
\affil*[1]{\orgname{A.K.Choudhury School of Information Technology},
\orgname{University of Calcutta},
\orgaddress{\city{Kolkata}, \state{West Bengal}, \country{India}}}

\author[2]{\fnm{Madhupa} \sur{Sanyal}}
\email{madhupa.sanyal@gmail.com}
\equalcont{These authors contributed equally to this work.}
\affil[2]{\orgname{Genpact}, \orgaddress{\city{Bangaluru}, \state{Karnataka}, \country{India}}}

\author[3]{\fnm{Ari} \sur{Kusumastuti}}
\email{arikusumastuti@gmail.com}
\equalcont{These authors contributed equally to this work.}
\affil[3]{\orgdiv{Department of Mathematics},
\orgname{Universitas Islam Negeri Maulana Malik Ibrahim Malang},
\orgaddress{\street{} \city{Malang}, \postcode{}, \state{Jakarta}, \country{Indonesia}}}

\author[3]{\fnm{Hrishav} \sur{Bakul Barua}}
\email{hrishav.barua@monash.edu}
\equalcont{These authors contributed equally to this work.}
\affil[3]{\orgdiv{},
\orgname{Monash University},
\orgaddress{\street{} \city{Malaysia and}, \postcode{}, \state{Melbourne}, \country{Australia}}}

\author[4]{\fnm{Kartick Chandra} \sur{Mondal} \orcidB}
\email{kartickjgec@gmail.com}
\affil[4]{\orgdiv{Department of Information Technology}, \orgname{Jadavpur University}, \orgaddress{\city{Kolkata}, \state{West Bengal}, \country{India}}}
\equalcont{These authors contributed equally to this work.}

\abstract{
Today's era is the digitized era. Managing such generated big data is an important factor for data scientists. Day by day, it increases the demand for big data storage systems. Different organizations are involved in providing storage-related services. They follow the different architectures or storage models for storing big data. In this survey paper, our target is to highlight such storage architectures  which provided by different renowned storage service providers. On an architectural basis, we divide the  big data storage systems into five parts, Distributed file systems (DFS), Clustered File Systems (CFS), Cloud Storage, Archive Storage, and Object Storage Systems (OSS). Also, we reveal a detailed architectural view of the big data storage systems provided by the different organizations under these parts.   
}

\keywords{Data management, Data storage, Object Storage, Cloud Storage, Cloud data storage, File Storage, Block Storage, Archive Storage, Big data storage}

\maketitle

\input{Introduction}

\input{Background}

\input{Storage}

\input{hybr}

\input{conclusion}
\backmatter

\section*{Statements \& Declarations}
\subsection*{Competing Interests and Funding}
The authors have no relevant financial or non-financial interests to disclose.

\subsection*{Conflict of Interest} The authors declare that they have no conflict of interest.

\bibliographystyle{sn-basic}
\bibliography{bibliography2}
\end{document}

%% file: Introduction.tex
\section{Introduction}
\label{sec:introduction}

\definecolor{darkgreen}{rgb}{0.0, 0.2, 0.13}
 	\definecolor{darkpastelgreen}{rgb}{0.01, 0.75, 0.24}

\begin{figure}[htb]
	\begin{center}
		\centering
		\includegraphics[width=13cm,height=6cm]{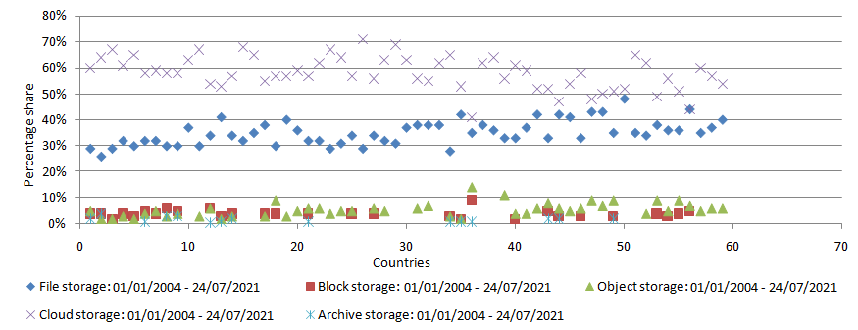}
		\caption{Country-wise plot of the popularity of storage systems as retrieved from Google Trends. (The readers are requested to view the online version of this colored image.)}
		\label{fig:bar}
	\end{center}	
\end{figure}

Figure \ref{fig:bar} shows the interest in Cloud, Object, File, Archive, and Block storages (y-axis, in percentage share) for some of the selected countries across the globe. 
We have selected the top 59 countries (x-axis) such as the United States, Australia, India, Canada, China, United Kingdom, Russia, and more on the basis of their search trend survey.
Figure \ref{fig:bar} clearly shows the popularity of Cloud storage over the other discussed storage systems.


Cloud storage is a service hosted by cloud providers such as Infrastructure-as-a-Service (IaaS), and Storage-as-a-Service (STaaS).
Data can be stored in logical pools built atop physical storage.
The physical storage is not actually hosted in one machine/server, rather, it can span multiple servers.
Physical storage is virtualized for efficient usage and management of huge amounts of data.
Generally, when we talk about cloud storage, we refer to object storage hosted by cloud providers across participating servers. Although other types of storage (such as block storage) are also being hosted by cloud providers as STaaS and used for Compute-as-a-Service, these are not as beneficial as object storage systems from the perspective of cloud services and Big un-structured data. 


The concept of cloud storage, centered around distributed resources, is suitable for implementation using object storage due to its features such as abstraction (from application and users), durability (through versioned copies), and fault-tolerance (via distribution and redundancy).
Object storage in the cloud is preferred due to its cost-effectiveness and reliability.

\textbf{Motivation and Uniqueness of this survey:}

The performance analysis between the storage systems always has been a debatable issue as present in \cite{69,73,88}.
In \cite{88}, performance analysis between cloud computing storages (distributed/scale out cluster file-systems) Compuverde \cite{88}, Gluster \cite{88} and Swift \cite{88} has been discussed.
In every case mentioned above, architectures have been explained in detail and main focus is on performance which is compared on the basis of read, write and delete operations.
The choice of cloud storages is purely based on allocation or data distribution technique i.e., Compuverde uses Multicasting for Distribution, Gluster \cite{88} and Swift \cite{88} uses Distributed Hash Algorithm (DHT).
However, emphasis on architectural characteristics comparison between these storages could have been an interesting point to discuss.



To maintain the increasing popularity of Cloud Storage systems and services based on Object Storage architectures, different storage service providers have launched their products. 
Each of these has different architectures based on their supporting properties. 
For example, the target of the Amazon Storage service provider is to support the storage data volume and telnet service request handling.
For this purpose, they launch different storage platforms: (Simple Storage Service (S3), Elastic Block Storage (EBS), and Elastic File System (EFS)).
The focus of the Microsoft vendors is to provide the storage platforms (Blob storage, Table storage, Queue storage, and File storage) based on the application demands.
Without the knowledge of architectures, storage service consumers will not be able to consume the needed storage system.
Therefore, our aim is to point out the architectures of diverse commercial cloud storage systems which will create awareness for the storage service consumers as well.

\begin{figure}[htb]    
\centering
\scalebox{0.55}{
	\begin{forest}
	for tree={rounded corners, draw, align=center, top color=white, bottom color=blue!20, edge+=->, l sep'+=1pt}, baseline, qtree
	[\color{brown}Organization of the article
		[\color{blue}Section \ref{sec:introduction}\\Introduction
		]
		[\color{blue}Section \ref{sec:background}\\Background Knowledge
		]
		[\color{blue}Section \ref{sec:store}\\Classification
			]
		[\color{blue}Section \ref{sec:concl}\\Conclusion
			]
	]
	\end{forest}
}

\caption{Organization of the article}
\label{fig:classi}
\end{figure}
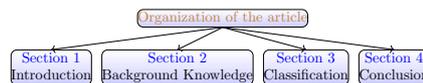

{\textbf{Organization of the survey:}}
This paper is organized as shown in Figure \ref{fig:classi}, it represents the section headings of this survey article.
This article begins with a background study (Section \ref{sec:background}) followed by the mechanism of object storage management.
In the next part, Section \ref{sec:store} discuss about the architecture of Big data storage systems included with distributed file system, clustered file system, cloud storage, archive storage, and object storage systems. 
At the last Section \ref{sec:concl}, concludes this article with referring the overview of this article. 

%% file: Background.tex
\section{\textbf{Background Knowledge}}
\label{sec:background}

For storing data as objects in locations and retrieving the same most efficiently, Object Storage Device or OSD \cite{67} performs the task of management of some services \cite{66}.
The services are discussed below in the following.

\begin{itemize}
	\item \textbf{Management of free space:} For efficient storage of objects, free space information must be maintained properly.
Also, proper data structures must be used to hold objects. 
These are two basic functions of any storage system. 
In a typical object storage system, we obtain discrete units called \qq{objects} by breaking the data into pieces. 
Then it is kept in a single repository, unlike a file storage system where it is kept as files in folders or as blocks on servers in a typical block storage system. 
Object storage system architecture resembles a flat structure where files (the actual data) are broken into pieces (objects) and spread out among hardware. 
All user objects are mapped to files (not like we see in a file storage system). 
Attributes of root objects, partition objects, and user objects are also stored as files but in a flat structure unlike hierarchical structure of a conventional file storage system. However, some of the stores like  Ceph Bluestore \footnote{ https://ceph.io/community/new-luminous-bluestore/} and IBM Cloud Object Storage \footnote{https://community.ibm.com/community/user/storage/blogs/praveen-viraraghavan1/2020/07/15/ibm-cloud-object-storage-cos-increases-performance} goes for block device rather than file system.

	\item \textbf{Security:} Security is another main functionality of OSD.
	As shown in the Figure \ref{fig:security}, in order to access the objects, clients or users must request for credential information from the security manager.
	The security manager is structurally composed of modules which maintains the credential information.
	This information is shared with an OSD via a capability key.
	Thus, the OSD check whether a user has requisite right to perform a particular operation.

	\begin{figure}[htb]
		\centering
		\includegraphics[width=13cm,height=3.5cm]{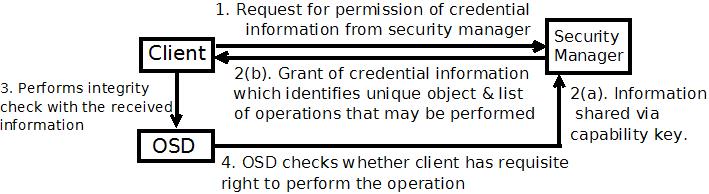}
		\caption{Security mechanism for storage and management of Object Data.}
		\label{fig:security}
	\end{figure}

	\item \textbf{Command Interpreter:} Command Interpreter is responsible for converting object command to a format that is used for storage system that lies underneath \cite{66}.
	Actually, it converts a command to a file system call.
	The stages of execution of this interpretation process (as given in Figure \ref{fig:command}), is explained in the following.

	\begin{figure}[htb]
		\centering
		\includegraphics[width=8.5cm,height=3cm]{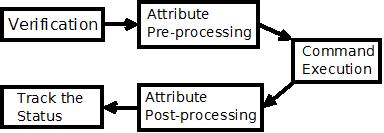}
		\caption{Stages in execution of Command Interpreter. (Object Data.)}
		\label{fig:command}
	\end{figure}

	\begin{itemize}
		\item \textbf{Verification:} At first, the Interpreter verifies whether the command has right to access and execute on the object.
		
		\item \textbf{Pre-processing:} Every command will either get or set the attribute of the target object.
		The pre-processing stage is to check the type of command from a group and decide whether to get or set attribute \cite{66}.
		
		\item \textbf{Execution:} Actual processing of command on the target object takes place in this stage.
		The object command is translated to storage-system specific system call.
		For example, if an OSD command is \qq{OSD WRITE} and the underlying storage system is a file, then the command is translated to \qq{write()} system call \cite{66}.
		
		\item \textbf{Post-processing:} This is the stage where final update process takes place and all attributes related to the command are changed.
		One simple example is during post-processing of \qq{write operation}, the timestamp of the file related to that object is updated.
	\end{itemize}
\end{itemize}

%% file: Storage.tex
\section{\textbf{Taxonomy of Big Data Storage System}}
\label{sec:store}

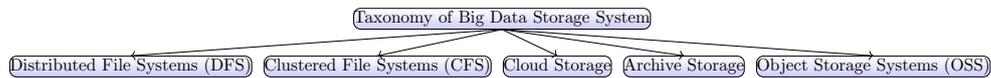
\begin{figure}[h]
\centering
\scalebox{0.65}{
	\begin{forest}
	for tree={
    rounded corners, draw, align=center, top color=white, bottom color=blue!20,
    edge+=->, l sep'+=1pt,	}, baseline, qtree
		[\color{black}Taxonomy of Big Data Storage System
			[\color{black}Distributed File Systems (DFS)]
                [\color{black}Clustered File Systems (CFS)]
			[\color{black}Cloud Storage]
			[\color{black}Archive Storage]
			[\color{black}Object Storage Systems (OSS)]
		]
	\end{forest}}
	\caption{Taxonomy of Big Data Storage System.}
	\label{fig:impl}
\end{figure}

\subsection{\textbf{Cloud Storage}}
\label{subsec:cloud}

As briefly introduced, the cloud storage provides a platform which abstracts the object storage architecture at back-end.
There are many cloud providers, but few utilize object-storage services.
The data can be accessed via graphical user interface or REST API \cite{78}.
Also various JAVA libraries can be utilized to access the objects in the cloud.
Amazon S3 \cite{42,43,69}, Mezeo Cloud storage \cite{2}, EMC Atmos \cite{73,93}, EMC ECS \cite{28,29,30} and Eucalyptus Walrus \cite{92} belong to this category (Table \ref{tab:cloud}).

\begin{table}[tbh]
{\small
	\caption{Summaries of some Cloud Storage type of Object-based storage.}
	\label{tab:cloud}
	\begin{tabular}{|p{3cm}|p{9.5cm}|}
	\hline
		
		\textbf{Name of object storage} & 
		\textbf{Summarized architectural components}
		\\\hline

		Amazon S3 \cite{42,43,69} & 
		\begin{minipage}{9.5cm} 
			Here Buckets are created which are containers of data.
			Every object is stored in a bucket.
			Object= Metadata + Object Data.
			The metadata comprises name and corresponding value that describes the object.
		\end{minipage}
		\\\hline

		Mezeo Cloud \cite{2} & 
		\begin{minipage}{9.5cm} 
			The Mezeo Cloud architecture comprises 3 layers: Top Layer (Requests are handled by this layer via API's and REST Web Services), Middle Layer (Provides services like allowing data to be stored in 1 or more files) and Lower Layer (It is the actual storage of data) and the cloud platform 
		\end{minipage}
		\\\hline

		EMC Atmos \cite{73,93} & 
		\begin{minipage}{9.5cm}
			EMC Atmos is a cloud platform by EMC Corporation which also gives support file-based services like Network File System (NFS) to access data.
			An important feature of Atmos is \qq{multi-tenancy}   which implies to the system being hosted in one or more user site which are accessed by a number of tenants or users.
		\end{minipage}\\\hline
		
		EMC ECS \cite{28,29,30} & 
		\begin{minipage}{9.5cm}
			Buckets are the actual containers of data, which are distributed across sites.
			Important feature is providing mechanism of geo-caching which identifies the location of the site requiring maximum access to data and caching there.
		\end{minipage}\\\hline
	
		Eucalyptus Walrus \cite{92} & 
		\begin{minipage}{9.5cm} 
			Walrus is the object storage service of Eucalyptus in which users store data in buckets and objects.
			Walrus provides an interface for manipulation of buckets.
		\end{minipage}\\\hline

    \end{tabular}
}
\end{table}

\subsubsection{\textbf{Amazon Simple Storage System (S3)}}
\begin{figure}[tbh]
	\centering
	\includegraphics[height=5cm,width=12cm]{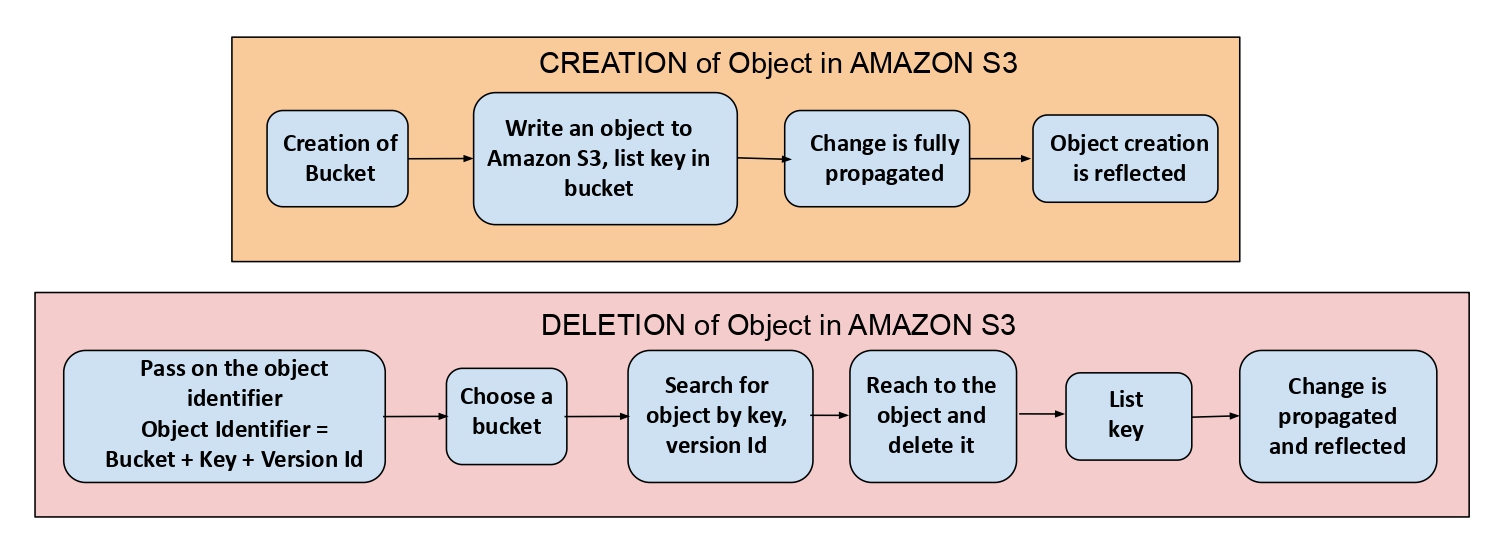}
	\caption{Operation of Amazon S3.}
	\label{fig:amazon}
\end{figure}

Amazon S3 is a simple implementation of Web Service \cite{43} which allows data storage in form of objects \cite{69}.
Buckets are created which are containers of data.
Proper authentication and access control mechanisms are also present for container and data management.
Components of Amazon S3 (as depicted in figure \ref{fig:amazon}) include the following.

\begin{itemize}
	\item \textbf{Bucket:} Every object is stored in a bucket.
	For example if the object is \qq{abc.jpg} which belongs to \qq{alpha} bucket, the image is addressable using URL http://alpha.s3.amazonaws.com/abc.jpg.
	Functions of bucket include defining the main storage and the corresponding path that leads to an object.
	Buckets also provide mechanism for access control.
	It can be configured to a specified region.
	When an object is added, a unique version ID is associated with it.

	\item \textbf{Object:} For Amazon S3, object is the basic unit of storage.
	Object = Metadata + Object Data.
	The Metadata comprises name and corresponding value that describes the object.
	Standard information like last data modified are present as default Metadata.
	An object is identified within a bucket by one key name and unique version ID.

	\item \textbf{Key:} Key name is unique for an object.
	Key is used to identify an object in basket.
	Bucket + key +version ID = identifier of the object.

	\item \textbf{Region:} Regions to store buckets are chosen to improve performance and minimize cost.
	For instance, some object data is of specific usage to only banks in Japan.
	Hence, data is stored in Asia Pacific Region to minimize network latency.
	Amazon S3 supports some region like Asia Pacific-Tokyo Region and EU-Frankfurt region.
	Objects stored in a region cannot move or exit from a region unless explicitly transferred.

	\item \textbf{REST API:} REST API uses standard HTTP request for storing, retrieval or  deleting bucket or object.
	We can use any tool that supports HTTP, or even browsers to access objects.

	\item \textbf{SOAP API:} New Amazon Features cannot be supported for SOAP API.
	But it can be still used with HTTPS.
\end{itemize}
Amazon S3 has been utilized to provide web services and in media and health care industry as well.

\subsubsection{\textbf{Mezeo Cloud Storage}}
Mezeo Cloud storage provides a platform to transform any traditional storage to scalable cloud storage.
Mezeo File Client ensures availability of data.
It provides solution to the current data management problems, by means of Mezeo Cloud and Mezeo File Solution \cite{2}.
Mezeo Cloud is deployed on LINUX based server and it follows a scalable Metadata storage.
The Metadata is assigned to every node in the cluster.
Hence, fast access to the data is possible without actually disturbing the storage.
Along with the object or file, the Metadata is created.
The information about the file is maintained in a distributed file catalog \cite{2}.
To ensure more availability, file catalog and Metadata should be separated from actual file location.
The Mezeo cloud architecture comprises 3 layers \cite{2} as follows.

\begin{itemize}
	\item \textbf{Top Layer:}
	Top Layer is the access layer.
	Requests can be handled by this layer via API's and REST Web Services.

	\item \textbf{Middle Layer:} 
	It provides services like allowing data to be stored in 1 or more files.
	It helps in data replication.
	The Storage connector provides advance features in addition to forming a bridge to the storage.
	Encryption of data can be done by Mezeo or at the storage level.
	Multiple connector modules are present which are supported by single cloud.
	They allow organizations to convert any type of storage to object store.

	\item \textbf{Lower Layer:}
	It is the actual storage of data.
	Data is stored in containers.

	\item \textbf{Mezeo Cloud Storage Platform:}
	Provides easier access to Mezeo Cloud Storage and provides a platform for development of mobile applications.
	API is built primarily on three concepts: \qq{Resource},  \qq{Representation}, \qq{Method}.
	The resources are actually data objects which are uniquely identified by URL.
	The object may be represented in a format such as XML/JSON.
	The representation is in standard form which is identifiable by storage.
	Methods include the creation or update or deletion of an object.
\end{itemize}

Mezeo Cloud storage provides various features which makes it popular \cite{2}.
The features are listed as follows.

\textbf{Authentication:} Mezeo Cloud incorporates authentication systems and they can utilize the advantages of CDMI (Cloud Data Management Interface) for proper functioning of the environment.
CDMI refers to the interface that can be used for management of data in the cloud.
The client can also utilize the interface to manage containers and data

\textbf{Multi-tenancy:} Each vendor uses separate storage providing physical separation of data.
Mezeo Cloud hosts various storage by means of consolidated interface.

\textbf{Flexible Integration Points:} Mezeo Interoperability API (IOP) provides access to applications used for storage systems like Amazon S3.
IOP provides the combination of backup facilities like Asigra
\footnote{http://www.asigra.com/cloud-backup-software}, Commvault \cite{48}.

\textbf{Rapid Scalability:} The stateless architecture of Mezeo Cloud ensures scalability and accounts, data can be stored dynamically.

\textbf{User Services:} The services provided include uploading files of any type, offering real time access to it.
Also, permission is granted only to specified role.
The access to shared data expires with time for security.
Mezeo Cloud storage can provide archival solutions.
It can also be utilized for web services.

\subsubsection{\textbf{EMC Atmos}}
EMC Atmos is a cloud storage service platform providing object based storage for storing Peta-bytes of data \cite{73}.
Combining the advantages of Atmos, third generation of storage called EMC Elastic cloud storage has been developed, which will be discussed later.

EMC Atmos is a cloud platform by EMC Corporation which provides unstructured data storage in the form of objects \cite{73}.
The objects can be accessed via web services like REST API.
It also gives support to file-based services like Network File System (NFS) to access data.

An important feature of Atmos is \qq{Multi-tenancy} which implies the system is hosted in one or more user sites which are accessed by a number of tenants or users.
These users may in turn provide services to sub-tenants to continue the chain.
Whatever the depth of the chain, each user is identified by user\_id.
The storage system is also highly scalable i.e., additional nodes may be added to the existing rack or new rack may be introduced to add new nodes.

Data protection in Atmos is maintained by a number of mechanisms.
For instance, a decision is given whether an acknowledgment will be sent after a successful \qq{write} is done.
This is known as \qq{Geo Mirror} technique of data protection  \cite{73}.
Another procedure is the decision to split the data into code and place data fragments at different distributed nodes.
Accumulation of all fragments will give the resultant data.
This technique is known as \qq{Geo Parity} techniques \cite{73}.
The policy manager is in charge of these data management.
EMC Atmos is widely used in health care industry.

\subsubsection{\textbf{EMC Elastic Cloud Storage (ECS)}}
EMC ECS provides software defined storage in accordance with the huge scaling requirement of the cloud \cite{30}.
ECS provides low cost access to public cloud with minimum risk.
The features include the following \cite{30}.

\begin{itemize}
	\item \textbf{Global Repository:}
	ECS urges the vendors to consolidate multiple storage system into a single global repository on which numerous applications run.
	It can be accessed anywhere.
	Buckets are the actual containers of data, which are distributed across sites.
	The architecture ensures proper replication and availability.
	There are certain consistency mechanisms provided by ECS which ensures consistency among replicas.
	Important feature is providing mechanism of Geo-caching which identifies the location of the site requiring maximum access to data and caching there.

	\item \textbf{Maintenance of Data Lake:} Presence of centralized data lake which allows data to be stored in original format by maintaining semantics.
	Storage and processing is managed by Hadoop Architecture 
	\footnote{http://hadoop.apache.org/}.

	\item \textbf{Multi-protocol API accessible Storage:}
	Allows organizations to store huge amount of unstructured data as objects.
	ECS supports standard API's like REST API.

	\item \textbf{Support for Multi-tenant Platform:} 
	Ideal to support multi-tenancy.
	Reporting is enabled to provide object count, object creation and storage utilization.
	Quota features ensure locking of buckets.
	The creation and deletion of bucket is transparent to administrators via REST API.

	\item \textbf{IoT Cloud Storage:}
	Provides storage to huge volume of data coming from intelligent devices.
	The end-to-end view of processing of storage system and Geo-caching scheme makes it suitable for IoT platforms as well as storage.

	\item \textbf{Cloud Scale Storage:}
	Refers to dynamic scaling feature to incorporate huge storage.
	It is ideal to support cloud and Big Data applications for storing mass data.
\end{itemize}

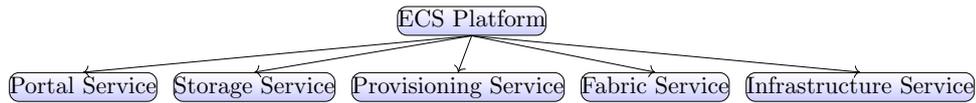
\begin{figure}[tbh]
\centering
\scalebox{0.9}{
	\begin{forest}
	for tree={rounded corners, draw, align=center, top color=white, bottom color=blue!20, edge+=->, l sep'+=1pt, }, baseline, qtree
	[\color{black} ECS Platform 
		[\color{black} Portal Service]
		[\color{black} Storage Service]
		[\color{black} Provisioning Service]
		[\color{black} Fabric Service]
		[\color{black} Infrastructure Service]
	]
	\end{forest}}
	\caption{ECS Platform Services.}
	\label{fig:ecs}
\end{figure}

ECS Cloud platform provides various software services \cite{29}.
As shown in Figure \ref{fig:ecs} for instance.
Portal Services provide interface to storage resources like GUI (browser-based; called portal) or REST API (which can be utilized to design own portal) or CLI (which functions similar to GUI).
Provisioning services are responsible for user administration, authentication, management of resources and multi-tenancy.
Fabric Service monitors the health of nodes, disk, cluster.
Infrastructure service provides LINUX OS on nodes and manages hardware tools.
Storage service layer forms the heart of ECS Cloud Storage platform \cite{29}.
The unstructured storage Engine (USE) \cite{29}, performs main functions like managing transaction and maintenance of object namespace.
USE performs all write operations on containers called chunks.
However, data that is written is not updated and is written in append-mode.
Thus, locking mechanism is not required.
For tracking object location, an index is maintained which records object name, chunk id and offset id.
Storage engine also ensures erasure coding and recovery for protection of data.
EMC ECS is used for archival purposes in health care industry particularly in partnership with EMC Isilon  \cite{28}.

\subsubsection{\textbf{Eucalyptus Walrus Object Storage}}
Walrus is the object storage service of Eucalyptus in which users store data in buckets \cite{92}.
Walrus provides an interface for manipulation of buckets and objects.
The function of buckets are \cite{92} as follows.

\begin{enumerate}
	\item Buckets are used to store and manage Eucalyptus Machine Images (EMI).
	An image is actually collection of various application software and system software which is uploaded to Eucalyptus cloud.
		
	\item They are also used to store and manage user data.
\end{enumerate}

During Eucalyptus installation, two types of accounts are created: Admin Account and User account.
Third party tools like S3 curl \cite{42} is used for interaction with Walrus.
For storing, retrieval of object or creation/deletion of bucket, users use these CLI or GUI interfaces provided by third party tools.
Even S3 buckets can be accessed as local directories.

Walrus uses ACL (Access Control List) for giving restrictions on users access to bucket/object.
Secret Key and Access Key are required to authenticate the user.
Once authenticated, permissions may be granted.

\subsection{\textbf{Archive Storage}}
\label{subsec:archive}

In archive type storage, only focus is management of huge volume of data for an indefinite period of time.
Since time-scale is invariant, hence the hardware setup and architecture is such that it will incur as much less cost as possible.
Its performance metrics is based on ease of scalability, retention efficiency and cost, replication mechanism, security policy and many others.
The two examples that have been discussed are: HCP \cite{84} and Storiant \cite{10} in Table \ref{tab:archive}.

\begin{table}[h]
\caption{Examples of Archive Storage type of Object-based storage.}
\label{tab:archive}
\centering
{\small

	\begin{tabular}{|p{3cm}|p{9.5cm}|}
	\hline
	
	\textbf{Object Storage System} & 
	\textbf{Summarized architectural components}
	\\\hline

	\begin{minipage}{3cm} Hitachi Content Platform (HCP) \cite{84} \end{minipage} &
	\begin{minipage}{9.5cm}
		Each object mainly comprises 2 components Fixed-content data which is the actual copy of the file contents and associated metadata which contains additional information like time of creation and Access Control List (ACL) defining set of permission given to users.
	\end{minipage}
    \\\hline

	Storiant \cite{10} & 
	\begin{minipage}{9.5cm}
		Storiant is one of the leaders in provider of cold data storage where data is being stored for a long period at a cost which is lower than the traditional system.
		Storiant developed object storage by consolidating various open source technologies like Open ZFS file, Cassandra NOSQL for storing object metadata and for programmable interface.
	\end{minipage}
	\\\hline
	\end{tabular}
}
\end{table}

\subsubsection{\textbf{Hitachi Data System (HDS) - Hitachi Content Platform (HCP)}}
The exponential growth pattern and complexity of handling voluminous data is a great concern.
HDS provides a solution to these problems through HCP.
HCP is a \qq{multipurpose distributed object-based storage system} \cite{84} that helps in the management of unstructured data by protecting and retrieving data over a single platform.
The features of HCP \cite{84} are as follows.

\textbf{Object Based Storage:} 
Unstructured data files are stored as an object.
Object is a container which contains file data as well as Metadata.
Each object is an independent unit which is the target for manipulation of data.
The Metadata can be used to perform functions like storage tiering or allocation of object.
Each object is therefore intelligent to understand the advance storage and this ensures proper distribution.
HCP Architecture ensures abstraction of object data.
Internally, data is present in object container.
Externally, they are represented as set of files in a directory or by URL.

\begin{figure}[tbh]
	\centering
	\includegraphics[scale=0.7]{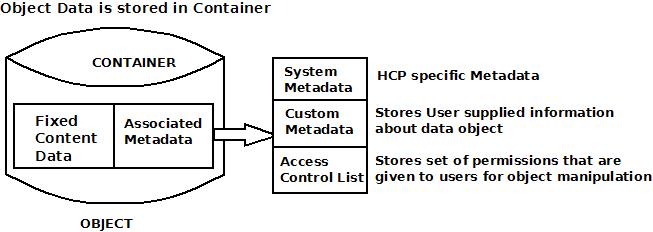}
	\caption{Object Structure in HCP Architecture.}
	\label{fig:hds}
\end{figure}

\textbf{Object Structure:}
Each object mainly comprises 2 components as shown in Figure \ref{fig:hds}: Fixed Content data and associated Metadata.
The metadata consists of System Metadata, Custom Metadata (optional) and Access Control List (ACL).
Fixed Content Data is an actual copy of the file contents.
After the file is stored, it becomes immutable.
If versioning is permitted, multiple copies may be present.
The Metadata describes content of data.
System Metadata may include the time the object was stored and modified (i.e., HCP specific Metadata) as well as Metadata like user id and group id.
Custom Metadata contains user supplied information about data object usually in XML Format.
ACL contains a set of permission given to users and a group of users to manipulate an object.

HCP also has open architecture that abstracts the data from technology enhancements.
This ensures proper archival where data can be accessed by users even after a long time period.
Hence, HCP provides storage as well as enables access via several interface like HTTP REST.
Hitachi Adaptable Modular Storage (HAMS) \cite{18} is introduction to extend the functionalities of HCP.
It is capable of handling more complex tasks mostly suitable in high-end storage system.
HCP has applications in health care industry for storing medical records.

\subsubsection{\textbf{Storiant Object Storage}}
Storiant is one of the leaders in provider of cold data storage.
Cold data storage actually refers to the mechanism where data is being stored for a long period at a cost which is lower than the storage cost of traditional system \cite{10}.
Storiant developed object storage by consolidating various open source technologies like Open ZFS file \cite{85} and Cassandra NOSQL \cite{77} for storing object Metadata and Swiftstack \cite{45,83,88} for programmable interface.
Here, object is stored in logical containers.
Open ZFS is an example of distributed file system which is highly scalable with features like replication, compression and data security \cite{85} running on LINUX operating systems.
Cassandra NOSQL is an example of Key-Value Store.
In Key-Value store, data is indexed on the basis of key and the value is associated with it.
ZFS is actually used for high reliability purposes.
Cassandra is used for proper storage.
Storiant uses MAID (Massive Array of Idle Disks) technology \cite{64} for physical storage purposes.
MAID technology comprises disk drives that are constantly spinning.
Whenever a drive is out of order, it shifts data to other disks.
MAID together with Open ZFS and Cassandra makes Storiant ideal for storing unstructured cold data \cite{10}.

\subsection{\textbf{Object Storage System (OSS)}}
\label{subsec:oss}
OSS came into picture with a promise to deliver the same good performance and efficiency as that of other cloud storage providers.
OSS is actually a software solution deployed by organizations within their enterprise or in platforms belonging to various cloud providers.
Various examples of OSS and their architectural components are listed in Table \ref{tab:oss}.

\begin{table}[ht!]
	\caption{Examples of Object Storage System.}
	\label{tab:oss}
	\centering
	{\small
	\begin{tabular}{|p{3cm}|p{9.5cm}|}
	\hline

	\textbf{Name of Object Storage} &
	\textbf{Summarized Architectural Components} 
	\\\hline
	
	Scality RING \cite{82} &
	\begin{minipage}{9.5cm}
		The ring forms the heart of the system.
		Two basic processing components are present i.e Access Layer to receive data request from application server and Storage Layer to provide interface with physical storage device.
	\end{minipage}
	\\\hline

	IBM Cleversafe \cite{15} &
	\begin{minipage}{9.5cm}
		Major components include Cleversafe Manager for performance monitoring, Cleversafe Accesser for providing interface for manipulation and encryption of data and Cleversafe Slicestor which is the storehouse of slices of data.
	\end{minipage}
	\\\hline

	DDN WOS \cite{14} &
	\begin{minipage}{9.5cm}
		Major components include WOS building blocks, managed by WOS Core Software and WOS Interface.
	\end{minipage}
	\\\hline

	Amplidata Himalaya \cite{8,44} &
	\begin{minipage}{9.5cm}
		Against the architecture followed by previous storage systems of Amplidata example Amplistor, Himalaya incorporates a second layer called Scaler layer which contains reverse proxy layer to ensure protection to storage pool.
	\end{minipage}
	\\\hline

	Cloudian Hyperstore \cite{16} &
	\begin{minipage}{9.5cm}
		The components of Cloudian HyperStore include HyperStore Geo Cluster for storage and  HyperStore service components like S3 Service, Hyperstore Service which manages Hyperstore storage system, Redis DBservice for managing user information
	\end{minipage}
	\\\hline

	Caringo Swarm \cite{89} &
	\begin{minipage}{9.5cm}
		It is an example of distributed scale out type object storage system that supports different API and allows file, block level access of data stored in Clusters.
		The Swarm Manager  is responsible for accepting instructions coming from Swarm Cluster.
	\end{minipage}
	\\\hline

	StorageGRID \cite{68} &
	\begin{minipage}{9.5cm}
		The unique feature of StorageGRID is the use of Dynamic Policy Engine that is used in place of containers.
		The Dynamic Policy determines the location and the physical storage device to store data, the degree of replication.
	\end{minipage}
	\\\hline

	Huawei OceanStor UDS \cite{32} &
	\begin{minipage}{9.5cm}
		Main feature is the use of Sea of Drives (SoD) Architecture which comprises Access Cluster which processes client requests  and Storage Cluster containing Smart Disks to store data. 
		Access Layer processes client requests and storage cluster.
	\end{minipage}
	\\\hline

	Swift \cite{45,83,88} &
	\begin{minipage}{9.5cm}
		Swift is an open source object storage system comprising of Swift Cluster and Process Layer and the ring.
		Cluster is an aggregation of nodes which run on Swift Server and Services.
		The swift server processes running in Swift Cluster are referred to as Swift Server Process Layer.
		The functionality of Ring is mapping of names of entities to their physical location
	\end{minipage}
	\\\hline

	GridBank \cite{49} &
	\begin{minipage}{9.5cm}
		Its components are GridBank File System (which ensures that the object storage utilizes any of media storage like tape or cloud as need), Data Management (concerned with data security , integrity, retention of data) and Metadatabase (management of metadata).
	\end{minipage}
	\\\hline

	Lattus \cite{46} &
	\begin{minipage}{9.5cm}
		Comprises Controller Nodes (Distributes Object Data), Storage Nodes (Storage of Data) and Access Nodes.
	\end{minipage}
	\\\hline
	\end{tabular}
}
\end{table}

\subsubsection{\textbf{Scality RING Object Storage System}}
The ring forms the heart of Scality Object Storage System \cite{82}.
It follows scale-out system storage and shared-nothing architecture.
It was designed primarily with three goal.
\begin{itemize}
	\item Store data in the order of exabytes.
	\item Provide security and enhance durability.
	\item Manage data efficiently.
\end{itemize}

The Distributed Ring Architecture is based on three principles:
\begin{enumerate}
	\item {\bf Divide and Conquer:}
	This principle implies balancing of load across the distributed access points.
	There is no single point of failure.
	\item {\bf Divide and Store:} 
	The method refers to the distributed storage of data where all components work independent of each other and possibly in a parallel manner.
	\item {\bf Divide and Serve:} 
	This principle describes the replication of data across nodes thereby increasing storage performance and reliability.
\end{enumerate}

\begin{figure}[htb]
	\centering
	\includegraphics[height=5cm,width=12cm]{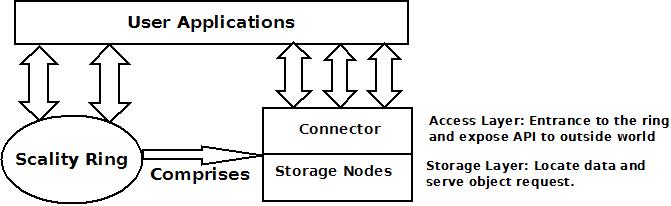}
	\caption{Scality RING Architecture.}
	\label{fig:sc}
\end{figure}

To provide high performance and scalability, Scality provides two basic processing components i.e., Access Layer and Storage Layer which are actually called Connectors and Storage Nodes \cite{82} as shown in Figure \ref{fig:sc}.

\textbf{Connector:}
The connector receives data request from application server and manipulates access to RING.
It serves as an entrance to the RING and exposes the API or implemented protocol to the outside world to map the data request to efficient routing mechanism for storing data.
Scality supports wide range of connectors including REST API and Scality FUSE.

\textbf{Storage Nodes:}
These are logical, virtual servers present in the RING and are distinctly separated from physical entity.
Main functions are read, write and retrieval of data.
Also, they provide interface and manages communication between the system and actual physical storage devices.

Every storage node is an independent LINUX/UNIX process.
Their main task is to locate data and serve object request.
Each server, by default, has 6 storage nodes.
These nodes are responsible for a separate segment of storage in global RING.
So, if there are 6 servers each having 6 nodes, 1 node is responsible for 1/36\textsuperscript{th}  of Key Space of Distributed Hash Table (DHT) that is present for locating object data.
Hence, if 1 node is non-functional, the corresponding 6 non-sequential portion of key space will be allocated to the remaining servers.
Scality is utilized for archival and analytics purposes by media broadcasters, in research labs as well as government services to manage big data in an efficient manner.

\subsubsection{\textbf{IBM Cleversafe Object Storage}}
\label{sec:clever}
\begin{figure}[htb]
	\centering
	\includegraphics[height=7cm,width=11.5cm]{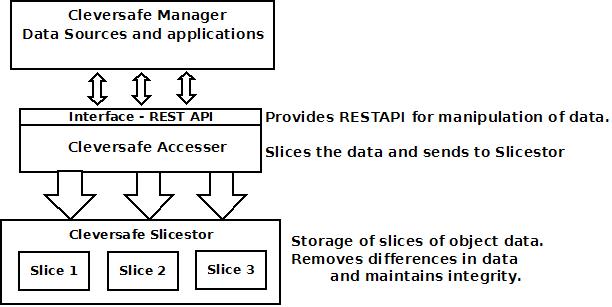}
	\caption{Cleversafe Software Solution.}
	\label{fig:clv}
\end{figure}

It is a software defined object storage solution that solves the problem of storing massive amount of data in order of petabytes.
It uses REST API for retrieval of large amount of object-based data that is stored in flat address container.
It used Information Dispersal Algorithm \cite{15} to allocate data that are split into slices.
The components \cite{15}, as shown in Figure \ref{fig:clv} are as follows.

\begin{itemize}
	\item \textbf{Cleversafe Manager:}
	It is a web-based manager for configuration, performance monitoring.
	\item \textbf{Cleversafe Accesser:}
	This provides REST API for manipulation of data.
	It also provides encryption of data.
	The data slices from the Accesser are sent to the third component i.e., Slicestor.
	\item \textbf{Cleversafe Slicestor:}
	It is the storage of slices and performs integration by removing differences in data.
\end{itemize}

Cleversafe is utilized immensely for analytics in R\&D Industry along with Hortonworks.
It is also used in health care industry for archival purposes.

\subsubsection{\textbf{DDN's Web Object Scalar (WOS)}}
DDN's chief importance is designing efficient storage system without making the structure more complicated.
WOS is an example of storage which strips down the architecture into basic units \cite{14}.
It comprises three components as shown in Figure \ref{fig:wos}: WOS Building Blocks, WOS Core Software and Interfaces.

\textbf{WOS Building Blocks:}
The unit of building block is WOS Storage node.
The storage nodes comprise 4U rack servers \cite{96} filled with SATA (Serial Advanced Technology Attachment) disk.
SATA is a new standard for connecting hard drive to computer system \cite{70}.
A WOS block by default has 3 nodes and also has flexible scalability.

\textbf{WOS Core Software:} 
Good storage needs to be managed by smart software.
WOS Core has efficient management functionality.
It even has \qq{self-healing} capacities \cite{14}, which reduces cost of maintenance.

\textbf{WOS Interface:} 
WOS provides decent choices of interface like S3 REST, file access interface.

WOS is designed as a true object storage platform following flat address space for storage.
The data is stored as contiguous blocks; hence, disk access is minimized and performance is enhanced.
It is mainly used as persistent storage in financial services and media and entertainment industry where replication factor, consistency are also very important.

\subsubsection{\textbf{Amplidata's Himalaya}}
Traditional object storage architecture describes objects which are self-contained with metadata.
Himalaya takes up the challenge by achieve unlimited scalability by attaching the Metadata to the physical layer of nodes \cite{8}.
Amplidata's previous object storage called Amplistore enhanced the speed of access to data by caching the object Metadata at controller nodes.
Hence, the entire storage pool need not be searched for the object.
Himalaya incorporates a second layer called Scaler layer.
The scaler layer contains reverse proxy layer to ensure protection to storage pool.

\begin{figure}[htb]
\begin{minipage}{0.2\textwidth}
	\centering
	\includegraphics[height=3.5cm,width=4cm]{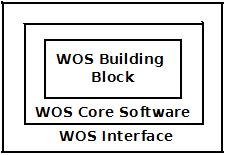}
	\caption{Components of WOS.}
	\label{fig:wos}
\end{minipage}
\hspace{10mm}
\begin{minipage}{0.70\textwidth}
	\centering
	\includegraphics[height=7.5cm,width=9.5cm]{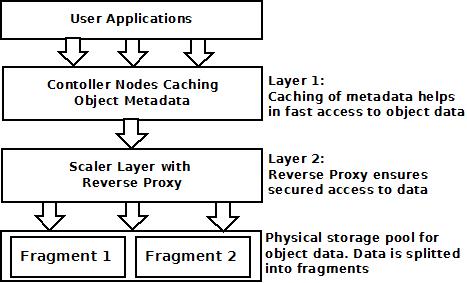}
	\caption{Himalaya Architectural Overview.}
	\label{fig:him}
\end{minipage}
\end{figure}

Amplidata's architecture has 2 components \cite{44} as shown in Figure \ref{fig:him}.
Scaler accessing devices and object store protected by erasure coding.
Erasure coding implies splitting of data into fragments, encoding with additional redundant bits and distributing across locations.
Scalers include highly scalable global name-space with access layer.
Reverse proxy also helps users to access and read data.
Himalaya's service is widely used in media and entertainment industry and as an archive in health-care industry.

\subsubsection{\textbf{Cloudian HyperStore}}
It is a multi-tenant supported object storage system \cite{16}, which supports Amazon S3 services.
It incorporates NoSQL storage layer for flexible and scalable storage of increasing upsurge of data.
It provides platform that deploys S3 services-compliant storage system.
HyperStore is again another perfect example of horizontally scalable object storage supporting increased storage of data by simply adding node in a flat address space.
It has got fully distributed, peer-to-peer architecture eliminating single point of failure.
Its various features \cite{16} include:

\begin{itemize}
	\item \textbf{Amazon S3 Compliant:} 
	All customers' existing S3 REST API based applications will work with Cloudian Service.
	S3 tools can also be used for developing HyperStore client applications.
	
	\item \textbf{Multi-tenancy:}
	Implies the  supporting multiple user access to single infrastructure in parallel.
	Each user's data is logically separated and permission needs to be granted explicitly for access.

	\item \textbf{Group Support:}
	A single HyperStore account can be shared by a group or an organization.
	Each group member supervised by an admin has a storage space allocated.
	
	\item \textbf{Quality of Service (QoS) Controls:}
	Cloudian system administrator can control usage rate per member of group or per user.
	Based on these quotas, reporting mechanisms ensure proper billing of group or individual users.

	\item \textbf{Access Control:}
	Read and write access are supported per bucket, per object.
	Objects are accessible to external environment via URL.
\end{itemize}

The components of Cloudian HyperStore include HyperStore GeoCluster for storage and services like S3 Services,  HyperStore Service, Redis DB 
\footnote{https://redis.io/}
Service and Cloudian Management Console (CMC).

\begin{figure}[tbh]
	\centering
	\includegraphics[height=10cm,width=12cm]{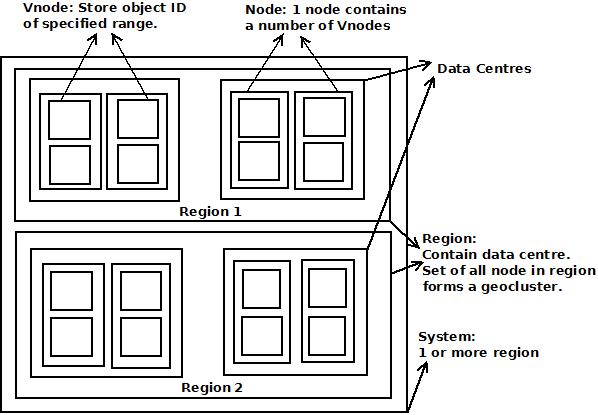}
	\caption{HyperStore GeoCluster.}
	\label{fig:hyp}
	\end{figure}

\textbf{Cloudian Hyperstore GeoCluster:}
As shown in Figure \ref{fig:hyp}, the entire hyperstore system can be deployed in several geographical locations.
Each region contains a number of data centers.
1 or more nodes are present in the data centers.
Each node contains a number of VNodes which store object for a particular id range.
Set of all nodes in a single region forms a GeoCluster.
Different regions store different datasets and there is no replication across regions.

\begin{figure}[htb]
	\centering
	\includegraphics[height=9cm,width=6cm]{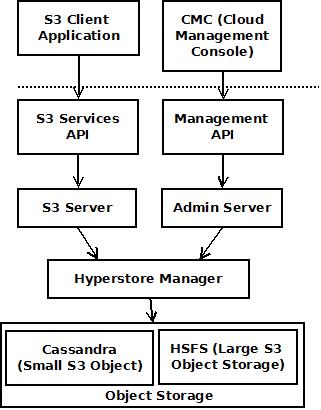}
	\caption{HyperStore Service Components.}
	\label{fig:hyp1}
\end{figure}

\textbf{Service Components:}
As shown in Figure \ref{fig:hyp1}, it supports several type of services as follows.
\begin{enumerate}
	\item \textbf{S3 Service:}
	Responsible for handling S3 REST requests coming from client applications.
	
	\item \textbf{HyperStore Service and HyperStore File System (HSFS):}
	HyperStore storage system uses hybrid storage system where Cassandra can be used for storing small S3 data objects and LINUX file system resting on Cassandra can be used for large S3 data objects.
	The portion of LINUX File system where object data is stored is called Cloudian HyperStore File System (HSFS).
	Casssandra defines key-space of object based storage.
	
	\item \textbf{Redis DB Services:}
	These are used to store user-level and group level QoS settings that is set by system administrator and to count user requests.
	
	\item \textbf{Cloudian Management Console (CMC):}
	It is a web-based UI (User Interface) for system administrator, group administrator, users for QoS control and manages billing plan, generates bill and views data stores as object.
\end{enumerate}


HyperStore is popularly used in media \& Entertainment industry as well as in R \& D labs for analytics purposes.

\subsubsection{\textbf{Caringo Swarm 7}}
Caringo Swarm 7 is an example of distributed scale out type object storage system that supports different API and allows file and block level access of data \cite{89}.
It is an object based software solution that can run on  hardware platform which is compatible with Linux or virtual machines.
New nodes can be added dynamically in an existing cluster as and when needed.
Swarm does not have a separate Metadata management.
Objects (i.e., Data + Metadata + Policies) are directly stored on disk.
This ensures block level access in addition to file system (NFS) and object API access methods (REST API).
Copy of index pointing to Metadata is stored in RAM for faster access.
Swarm enables different gateways like Cloud Scalar and File Scalar, each of which provides platforms to traditional interfaces as well as S3 API.
These interfaces are selected by users.

Swarm Manager \cite{89}, is responsible for accepting instructions coming from Swarm Cluster.
It also schedules the proper resources against a particular cluster.
The failure of Swarm Manager leads to the failure of cluster operations.
The actual processing is done by primary Swarm Manager.
The Secondary Swarm manager forwards the same to primary manager.
Swarm is used for storing huge volume of unstructured data on platform for Big Data Analysis.

\subsubsection{\textbf{NetApp StorageGRID Webscale}}
\begin{figure}[htb]
	\centering
	\includegraphics[width=12cm,height=6.5cm]{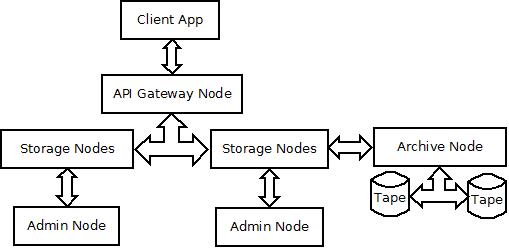}
	\caption{NetApp StorageGRID Webscale Architecture.}
	\label{fig:net}
\end{figure}

StorageGRID Webscale is a storage platform built for object-based data storage technique \cite{68}.
It has massive scalability to store billions of objects in distributed nodes running under single global namespace.
The unique feature of StorageGRID is the use of \qq{Dynamic Policy Engine} \cite{68} that is used in place of containers (a unique manager of data for object storages).
The Dynamic Policy determines the location and the physical storage device to store data, the degree of replication (number of copies of data to be stored) and the Metadata stores information about the location of data and the number of fields.
StorageGRID also provides S3 API support which is the standard API for object storage.
CDMI support is also given.
The architecture of StorageGRID Webscale as shown in Figure \ref{fig:net}.
It comprises 4 nodes.

\begin{itemize}
	\item \textbf{Admin Nodes:}
	It is responsible for services like configuration, audit, logging and other management services.

	\item \textbf{Storage Node:}
	It controls object storage.
	It also ensures proper replication of object.
	
	\item \textbf{API Gateway Node:}
	It provides an API through which applications connect to StorageGRID.
	
	\item \textbf{Archive Node:}
	It provides interface to physical archival medium like tapes \cite{68}.
\end{itemize}

\subsubsection{\textbf{Huawei OceanStor UDS}}
Based on big data market demand, Huawei released UDS, a massive storage system that follows scale out architecture and appropriate data protection mechanism \cite{32}.
UDS comprises Access Nodes (A Node) and Distributed Nodes (UDSNs).
A Nodes are primarily used for data scheduling i.e., assignment of data requests to particular UDSNs.
UDSNs are used for data storage.
The architecture followed is known as \qq{Sea of Drives (SoD) Architecture}.

\begin{figure}[htb]
	\centering
	\includegraphics[width=12cm,height=7cm]{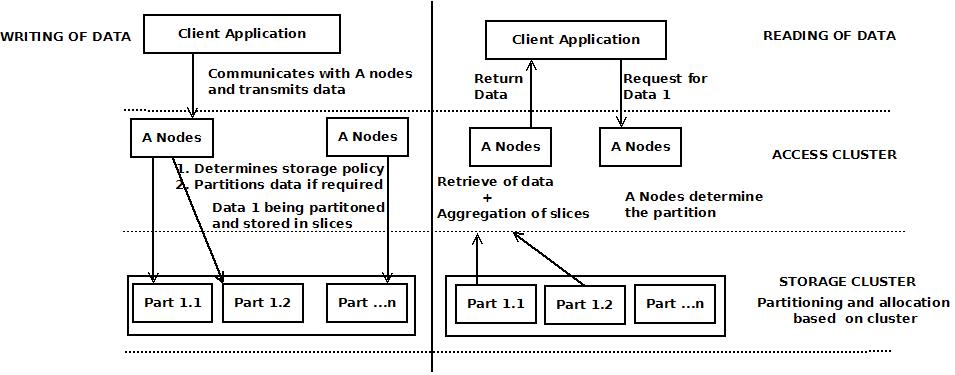}
	\caption{UDS Sea of Drives(SoD) Architecture and I/O Process.}
	\label{fig:uds}
\end{figure}

\textbf{Sea of Drives (SoD) Architecture:}
Sea of Drives is an architecture suited for manipulation of huge amount of unstructured data that is used primarily for more read operations than write.
SoD architecture comprises Access Cluster and Storage Cluster as shown in Figure \ref{fig:uds}.

\begin{itemize}
	\item \textbf{Access Cluster:}
	Access Layer comprises A Nodes that processes client requests.
	It controls routing of a client request to a proper UDSN.
	A Nodes are accumulated to form clusters.
	When the number of requests increase, the nodes can be added dynamically to reduce traffic.

	\item \textbf{Storage Cluster:}
	Storage Cluster comprises UDSNs which contain Smart Disks.
	Smart Disk is the basic unit of physical storage.
	It contains disk drive together with CPU, memory and Ethernet Port.
	Each disk is designated with dedicated IP address and connected to switch and other disks.
	UDS capacity is increased by adding more disks.

\end{itemize}

\textbf{I/O Process:}
I/O process implies writing of object data and reading of object data.
Following steps are performed for writing data takes place in the Figure \ref{fig:uds}.

\begin{itemize}
	\item A client communicates with an A-node and transmits data to the A Node.
	\item A Node determines data storage policy (which involves determining multiple copy-for each data, use of erasure Code mechanism for data protection and so on).
	\item If data size is large, A Node splits the data into partitions.
	\item Then, A Node writes data slices into storage device based on DHT (Distributed Hash Table).
\end{itemize}

As shown in Figure \ref{fig:uds}, steps for reading data are as follows.
\begin{itemize}
	\item A client sets up connection with A Node.
	Specifies the data request.
	\item A Node determines the partition and retrieves address of the smart disk.
	\item A Nodes aggregate the data slices.
\end{itemize}

\subsubsection{\textbf{OpenStack Swift}}
\begin{figure}[tbh]
	\centering
	\includegraphics[width=12cm,height=11cm]{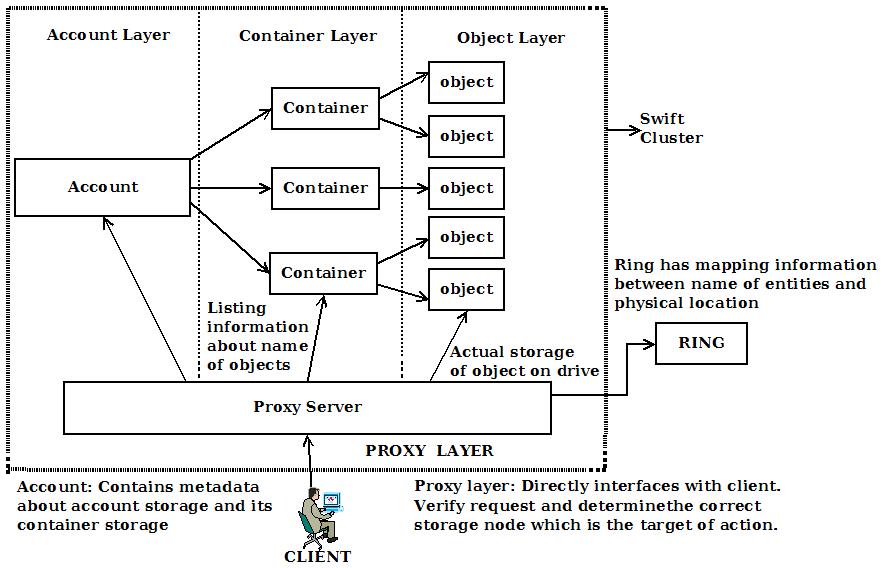}
	\caption{SwiftStack Architecture.}
	\label{fig:swiftarch}
\end{figure}

Swift \cite{83,88} is an open-source object storage system that stores unstructured data \cite{45} like files or videos in large quantities.
Object is the unit of data storage and data is replicated across various systems.
The applications access data via REST API.
An architectural overview of SwiftStack components and processes as shown in Figure \ref{fig:swiftarch} includes the following.

\begin{itemize}
	\item \textbf{Swift Cluster:}
	It follows the distributed object storage system concept.
	It is an aggregation of nodes which run on Swift Server and Services.

	\item \textbf{Swift Server Processes:}
	Four Swift Server processes are proxy, account, container and object.

	\begin{itemize}
		\item \textbf{Proxy:}
		Responsible for connecting the rest of Swift architecture.
		For each request \cite{83,74}, the proxy looks up for the location or account or object in the ring.
		All these components will be discussed in the following.

		\item \textbf{Object:}
		It is a simple BLOB storage system that involves storing, removing and retrieving objects on a local device \cite{83}.
		Objects are stored and identified by using path name.

		\item \textbf{Container:}
		Primary function is to list the objects \cite{83,74}.
		It does not know the location of the object.

		\item \textbf{Account:}
		It lists the containers that are present \cite{83,74}.
		Account should not be confused with user account.
		All accounts and containers are considered as places of storage.
	\end{itemize}

	\item \textbf{Swift Server Process Layer:}
	The swift server processes which are running in Swift Cluster are referred to as Swift Server Process Layer \cite{74}.
	There are four process layers.
	\begin{itemize}
		\item \textbf{Proxy Layer:}
		Directly interfaces with the client \cite{45}.
		It is the first layer to handle API.
		It uses standard HTTP request and response.

		\item \textbf{Account Layer:}
		Requests regarding Metadata \cite{45} of individual account or container are handled here.

		\item \textbf{Container Layer:}
		Requests regarding Metadata of object \cite{45}  are handled here.

		\item \textbf{Object Layer:}
		Responsible for actual storage of data on drive \cite{45}.
		Object are stored in binary file using path name.
	\end{itemize}

	\item \textbf{Ring:}
	Basic functionality of Ring is mapping of names of entities to their physical location  \cite{47}.
	Separate ring is maintained for account and container. 
	For one object, one ring is available.
	Whenever a component needs to perform an operation on an object or container or account, it needs to consult with the ring to determine the location in the cluster.
\end{itemize}

\begin{figure}[htb]
	\centering
	\includegraphics[height=2.5cm,width=12cm]{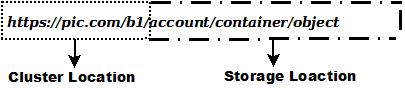}
	\caption{A Storage URL in Swift for an object.}
	\label{fig:swifturl}
\end{figure}

Every data or object in SwiftStack clusters is represented by means of URL.
A Storage URL in Swift is shown in the Figure \ref{fig:swifturl}.

The URL comprises two parts; Cluster Location and Storage Location.
\begin{itemize}
	\item \textbf{Cluster Location:}
	It determines target cluster address where the request will be sent.
	Say the storage URL is https://pic.com/b1/account/container/object.
	\qq{https://pic.com/b1} portion indicates the cluster location within which the object storage must be identified.

	\item \textbf{Storage Location:}
	It defines where request is to be processed within a cluster .
	A storage location can be defined in terms of account or container or object.

	\begin{itemize}
		\item \textbf{/account:}
		It is a storage area which lists the Metadata about the account and its corresponding containers.
		
		\item \textbf{/account/container:}
		Metadata about container and corresponding objects that are stored.
		
		\item \textbf{/account/container/object:}
		Storage comprising object \(data\) and its Metadata.
	\end{itemize}
\end{itemize}

\subsubsection{\textbf{Tarmin GridBank}}
Tarmin GridBank provides object storage solution for proper storage and management of object as data \cite{49}.
GridBank is mostly popular as a data repository in Oil \& Gas Industry.
Three components of GridBank actually utilizes the advantages of Data Defined Storage (consolidating data into an integrated storage where the users and application actually accesses the Metadata repository, thereby maintaining full abstraction about data).
The main components of Tarmin Data Defined Storage are GridBank File System (GBFS), GridBank Data Management, GridBank Metadatabase.
There are additional components too.

\begin{itemize}
	\item \textbf{GridBank File System (GBFS):}
	The file system ensures that the object storage utilizes any of media storage like tape or cloud as per organizational demand \cite{49}.
	The File system integrates all such storage and externally it gives a view of consolidated distributed data storage via the global namespace. 
	There is another utility, compresses huge volumes of data with the help of optimization layer.
	GBFS also has mechanism for removing duplication.

	\item \textbf{GridBank Data Management:}
	This layer is concerned with data security (encryption), integrity, retention of data and replication of data across various sites \cite{49}.

	\item \textbf{GridBank Meta Database:}
	Concerned with management of Metadata.
	It is actually a Metadata repository that contains 3 kinds of Metadata indexing from raw data.
	\begin{enumerate}
		\item File property and attribute
		\item Full content Metadata
		\item Arbitrary Key-Value Metadata
	\end{enumerate}
	The distributed store integrates the data coming form different datasets and forms consolidated Metadata access to data \cite{49}.
	
	\item \textbf{Policy Engine:}
	That is concerned with automation of migration of data and management of data coming from all sources.

	\item \textbf{Global Namespace:}
	Global namespace is maintained which provides a view to client with various protocols.

	\item \textbf{Multiple Protocol Access:}
	Data can be accessed via HTTP and FTP.
	GridBank also supports REST API for management of data.
\end{itemize}
GridBank is used extensively as storage medium of data in Oil \& Gas Industry.

\subsubsection{\textbf{Quantum Lattus}}
Existing traditional as well as current technologies face the issue of archiving data, as well analyzing or reusing it in future with a major focus on its availability \cite{46}.
Lattus is an example of scale-out architecture which eradicates this issue by providing online storage and its inherent nature of high scalability and durability.
It actually comprises three components \cite{46}.

\textbf{Controller Nodes:}
It stores data as object and distributes them into storage nodes.
The data can be accessed via REST API.

\textbf{Storage Nodes:}
It provides scalable object storage and can automatically check data integrity and perform bit error correction.

\textbf{Access Nodes:}
Access nodes improve performance by the mechanism of \qq{in-memory caching and disk caching} \cite{46}.
In-memory cache refers to keeping frequently accessed data in memory to reduce the latency of user query.
Disk Cache is a process which aims to reduce access time of hard disk.
It may be a part of hard disk or a portion of RAM which holds recently used data.
Lattus provides the following advantageous features which make it a strong contender for object storage \cite{46}.

\begin{itemize}
	\item \textbf{Scalability:}
	Scalability is achieved to thousands of petabytes of data without migration to other storages.
	Even automatic replication process takes place and proper distribution of data is ensured.

	\item \textbf{Self Healing:}
	There are error correcting algorithms which ensure that data is present even when there is a failure in node or device.
	Lattus continuously performs bit checking for errors and rectifies them.
	When the drives are replaced, data is again redistributed across network to ensure undisturbed availability of data.

	\item \textbf{Reduces Cost:}
	Since a great amount of replicated data is always available, backup cost is minimized.
\end{itemize}

%% file: hybr.tex
\subsection{\textbf{Distributed File System (DFS)}}
\label{subsec:dfs}
Distributed File System creates an illusion of a common file system by distributing the data file in different file servers, located in different locations \cite{dfs2015}.
These file servers are connected with each other via the internet and they are accessed via remote machine call.
This is the basic structure where fault tolerance and scalability are resolved.
Modern technologies are involved in distributed file systems for storing big data.
The architecture of ceph \cite{94,95} distributed file system is described in given below.


\subsubsection{\textbf{RedHat Ceph}}
\begin{figure}[htb]
	\centering
	\includegraphics[width=8cm,height=3cm]{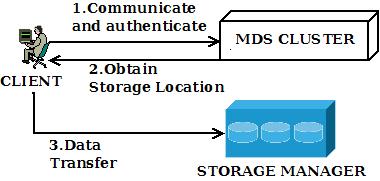}
	\caption{RedHat Ceph Storage System.}
	\label{fig:ceph}
 \end{figure}

Ceph is an example of distributed file system which offers high performance and scalability.
A unique feature of Ceph is separation of data and Metadata operations \cite{94}.
Metadata Cluster is only concerned with opening and renaming file (Metadata operation). 
Storage cluster is concerned with retrieval of object data.
There is a complete removal of overhead  caused by maintaining allocation list.
Instead, it forms an allocation function called 'CRUSH' \cite{94} which determines the location of the object in its partition at an OSD.
Hence, in order to access object, every time functional computation is performed by client instead of lookup for object and its location.
As the Metadata cluster need not be communicated for lookup, load is reduced.
Ceph has got the following architectural components \cite{95} as shown in the Figure \ref{fig:ceph}.
Major components are as follows.
\begin{itemize}
	\item \textbf{Metadata Server (MDS):}
	The main functions of MDS are controlling Metadata operation like file management, ensuring security, and proper authentication.

	\item \textbf{Client Interface:}
	Provides API to manage storage system.
	It also provides the platform between applications and storage system.
	Performs allocation mechanism called CRUSH to locate the object slice.

	\item \textbf{Storage Manager:}
	It manages Object Storage, allocation of free spaces in OSD and scheduling management in each OSD.
\end{itemize}

When application wants to access data, it first communicates with MDS cluster and then authentication takes place.
If authenticated, computation of object storage location is performed.
Then, data transfer takes place directly with OSDs after consulting Storage manager.

\subsection{\textbf{Clustered File System (CFS)}}
\label{subsec:cfs}
In clustered file system, a common file is distributed among multiple storage servers and all the servers are treated as a single file system.
Here, storage space is shared among the servers, which increases the system performance with respect to better utilization of storage space. 
OASIS \cite{76,87}, Panasas \cite{81} and Lustre File System \cite{61} are discussed below as CFS.


\subsubsection{\textbf{OASIS}}
\begin{figure}[htb]
	\centering
	\includegraphics[width=10cm,height=4.5cm]{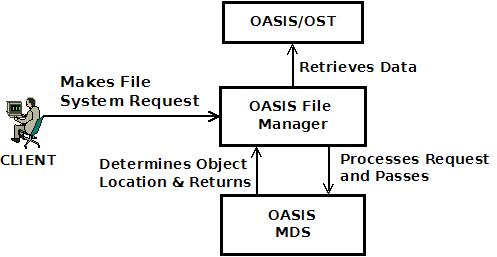}
	\caption{Flowchart of Functioning of OASIS.}
	\label{fig:oasis}
\end{figure}

It is an example of clustered file system utilizing OSD in a file system environment \cite{76}.
OASIS is designed with the aim of attaining a highly scalable structure.
The functioning of OASIS is shown in Figure \ref{fig:oasis}.
OASIS uses OSD as their storage device, and this also provides object based interface to the file data.
Each OSD is responsible for the optimization of storage and automatic storing of objects.
The Metadata cluster is responsible for global namespace management of object data \cite{76}.
It also helps in file to object mapping as well as maintains authentication and security of data.
There are primarily three components \cite{76}.

\begin{itemize}
	\item \textbf{OASIS OSD Target (OASIS/OST):}
	It is concerned with the management of object data in an OSD command interface.
	OST functions as an independent storage device which itself contains processor, RAM (Random Access Memory), the Gigabit Ethernet network interface and disks.
	OST actually comprises OSD manager and storage device driver.
	OSD manager implements T10 OSD storage protocol (T10 actually standardizes storage interfaces.
	It can be seen in reference \cite{87}) and uses SCSI transport protocol for manipulation of object data.
	Storage device driver is for supporting data storage in the file systems.

	\item \textbf{OASIS Metadata Server:}
	Concerned with operations like file creation, lookup and attribute assignment.
	It also is responsible for sharing file data.

	\item \textbf{OASIS/FS File System:}
	It provides POSIX API platform to client and is built on OASIS MDS and OST.
	It is the gateway to OASIS file and directory.
	For accessing data directly communicates with OSD.
\end{itemize}

\hspace{0.2cm}

%
%
%
%
%
%

\subsubsection{\textbf{Panasas}}
\begin{figure}[tbh]
	\centering
	\includegraphics[width=8cm,height=4cm]{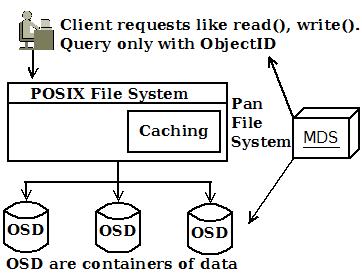}
	\caption{Panasas Architecture.}
	\label{fig:panasas}
 \end{figure}

Panasas Storage Cluster architecture is based on the concept of separation of Metadata from actual path leading to data \cite{81}.
Panasas storage devices are actually OSD, with which the client directly communicates.
For proper distribution of data, Panasas Active File System (Pan FS) stripes the data into number of objects and one object is allocated per OSD.
There are actually 5 components of Panasas as shown in Figure \ref{fig:panasas}.

\begin{enumerate}
	\item \textbf{Object:}
	Object is the unit of storage.
	Each object contains self-containing information which helps in proper management of data all by itself.
	Object, together with its attributes communicates with storage system regarding management of data.
	This is in contrast to a file system where the storage system monitors all the attributes of data.

	\item \textbf{OSD:}
	These are intelligent devices which structurally comprises disk drives, RAM, processor and provides three basic functionalities.

	\begin{itemize}
		\item \textbf{Data Storage:}
		Object data are present along tracks and sectors on storage devices.
		Clients request object data by querying with object ID, and offset that determines length of data requested.
		No block level access is permitted.

		\item \textbf{Intelligent Layout:}
		Provides a mechanism for efficient access to data.
		For example, since object Metadata states the length of data to be written, contiguous blocks can be determined.
		By the help of write-behind cache, huge amount of data is written in less number of passes.

		\item \textbf{Object Management:}
		OSD manages Metadata that is associated with the object.
		Hence, it reduces the load on MDS. 
	\end{itemize}

	\item \textbf{Panasas File System (Pan FS):}
	For reading and writing object from OSD, PanFS kernel module is loaded on the client.
	There are mainly 4 functions.

	\begin{itemize}
		\item \textbf{POSIX File System Interface:}
		Pan FS provides a POSIX interface to applications performing file operations (\qq{read()} and \qq{write()}).

		\item \textbf{Caching:}
		There are 3 kinds of caching facilities in PanFS.
		The first one provides caching in computation node for incoming data.
		The second one is write-data caching that  accumulates a stipulated number of writes for efficient transmission of data.
		Third caching allows clients to issue secured commands to access data on OSD.

		\item \textbf{Stripping:}
		Strips a file across multiple OSD where one object is allocated to one OSD.

		\item \textbf{iSCSI:}
		Uses iSCSI protocol to support OSD SCSI commands and transmit data payload over Ethernet. 
	\end{itemize}

	\item \textbf{MDS:}
	It is the bridge between clients and objects.
	MDS determines the map containing the outline of each object and thereby enables clients to access objects directly.
	It is concerned with the management of file stripping, ensuring secured access to object data.
	Secured access is done with the help of \qq{capability}, which is a secure token determining the client's authorization and token encoding object ID.
	Expiration of capability leads to rejection in client access.
	MDS performs reconstruction of lost data and cache consistency (notifies the client when changes in cache file are going to change the client's cache file).

	\item \textbf{Network Fabric:}
	Network provides efficient communication between OSD, MDS, PanFS and uses TCP/IP as protocol over Ethernet.
	iSCSI protocol is used for propagation of command and data from/to OSD.
	It has a facility of RPC (Remote Procedure Call) to provide fast communication between client and MDS.
\end{enumerate}

\subsubsection{\textbf{Lustre}}
\begin{figure}[tbh]
	\centering
	\includegraphics[width=10cm,height=4.5cm]{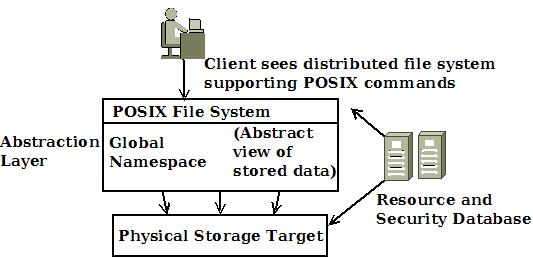}
	\caption{Lustre Architecture.}
	\label{fig:lustre}
\end{figure}

It is an example of open-source parallel distributed cluster file system using OSD as its storage device \cite{61}. It is generally used for large-scale computing.
It comprises the following components as shown in Figure \ref{fig:lustre}.


\begin{enumerate}
	\item \textbf{Client:}
	Absolute abstraction regarding the data storage is maintained from the client.
	The client sees a distributed file system supporting standard POSIX commands.
	A global namespace provides an abstracted view of available stored data.
	However, special applications can directly communicate with OSD without intervention of client.

	\item \textbf{Metadata Control System:}
	Concerned with management of file system Metadata as well as directly communicates with storage target for information about Metadata.

	\item \textbf{Storage Target:}
	It is the physical storage location of object.
	Objects can be an entire file, or strips of file data.
	The files are represented by container objects on MDS and data object in storage target.

	\item \textbf{Resource and Security Database:}
	Concerned with resource management and security services in databases.
\end{enumerate}

%% file: conclusion.tex
\section{Conclusion}
\label{sec:concl}
The big data storage system is a service model where internet technologies are used to store big data.
The data storage servers are placed in offsite locations and this data is accessed via internet service. 
Fault tolerance, accessibility, reliability, availability, scalability, and data security, all are managed by the big data storage system.
For big data managing purposes, different data storage service providers consider different approaches, such as distributed file system, clustered file system, cloud storage, archive storage, and hybrid storage solution.  

In this survey article, our aim is to present a brief overview of the architectural details of different big data storage systems.
These big data storage systems are developed by corporate organizations from a business perspective.
Different storage service provider has different intention to design storage system.
Some service provider serves as the data dump station and some support the data analytic-related job.
Therefore, the architectures are also varied from one to another. 
These architectural variations are manipulated through the detailed discussion of the architecture of different storage systems. 